\documentclass[12pt, letterpage]{article}
\usepackage{amssymb,latexsym}
\usepackage{natbib}
\usepackage[unicode=true, bookmarks=true, bookmarksnumbered=false, bookmarksopen=false, breaklinks=true, pdfborder={0 0 0},backref=false]{hyperref}
\hypersetup{
  colorlinks=true,
  allcolors=blue
}

\usepackage{amsthm}
\usepackage{tikz}
\usetikzlibrary{arrows.meta,positioning,calc,decorations.pathreplacing}

\usepackage{booktabs}

\usepackage{appendix}
\usepackage{amsmath}
\usepackage{bbold}
\usepackage{graphicx}
\usepackage{cases}
\usepackage{float}
\usepackage{subcaption}
\usepackage{multirow}
\usepackage{xcolor}
\usepackage{indentfirst,geometry,setspace,fancyhdr,sectsty}

\usepackage{hyperref}

\usepackage[colorinlistoftodos]{todonotes}
\allowdisplaybreaks

\newcommand{\jb}[1]{{\color{black} #1}}
\newcommand{\ao}[1]{{\color{black} #1}}
\newcommand{\cl}[1]{{\color{black}#1}}
\usepackage{marginnote}
\usepackage{marginfix}
\newtheorem{lemma}{Lemma}
\newtheorem{proposition}{Proposition}
\newtheorem{definition}{Definition}
\newtheorem{remark}{Remark}

\def\Pr{\hbox{Pr}}

\begin{document}
\title{\ao{\textbf{Moral AI and the Erosion of Its Own Demand}}}
\author{Wallice Ao\thanks{wallice.ao@gmail.com.} and Jiabin Wu\thanks{Department of Economics, University of Oregon, jwu5@uoregon.edu.}}

\maketitle

\begin{abstract}
\ao{We study the evolution of human preference frequencies and AI adoption when individuals can choose between a moral AI that rules out defection and an unconstrained AI that preserves discretion. By allowing users to commit to cooperation, moral AI can sustain cooperation even when cooperative preferences are too scarce to do so on their own. But this success has a cost. By turning otherwise dormant cooperative preferences into committed behavior, moral AI exposes its users to exploitation by selfish individuals, causing the population share of conditional cooperators to decline under evolutionary selection. As conditional cooperators become less prevalent, equilibrium adoption eventually converges to zero. Moral AI can therefore be self-liquidating. For initial shares between the moral-AI adoption cutoff and the threshold for cooperation without AI, moral AI generates cooperation while lowering the limiting share of conditional cooperators. Adoption ultimately vanishes, leaving a lower share of the preference type that initially sustained demand for commitment.}
\end{abstract}

\newpage

\section{Introduction}

\ao{
As AI takes on economic and social decisions, an important question is which forms of AI can survive competition.} Some systems may be designed with strong moral constraints to rule out harmful or exploitative actions, while others retain greater discretion.\footnote{Major AI developers already differ in how they restrict model behavior. OpenAI's \href{https://model-spec.openai.com/}{Model Spec} imposes platform-level rules that developers and users cannot override, while leaving lower-level behavior customizable. Anthropic's \href{https://www.anthropic.com/constitution}{Claude Constitution} similarly combines non-negotiable ``hard constraints'' with other behaviors that operators and users can adjust. Google's \href{https://ai.google.dev/gemini-api/docs/safety-settings}{Gemini API} allows developers to vary filtering thresholds for harassment, hate speech, sexual content, and dangerous content, while always blocking certain core harms. xAI's \href{https://x.ai/legal/acceptable-use-policy}{Acceptable Use Policy} instead emphasizes user control subject to legal, safety, and other guardrails. 
} \citet{hendrycks2023natural} argues that the latter may have a competitive advantage because they can behave responsibly when useful without giving up other profitable actions. This raises the possibility that competition may gradually select against constraints intended to limit harmful behavior.

The conclusion is less immediate once AI systems are used in strategic interactions. \jb{A constraint does not only remove actions; it can also change how others respond.} 
As AI agents become more capable, individuals and firms are likely to delegate an increasing range of consequential choices to them, making AI-mediated strategic interactions a routine feature of economic and social life. In these settings, restrictions built into an AI can serve as commitment devices. Commitment itself is not new. People have long limited their future discretion by signing contracts that penalize deviation, posting bonds or collateral that are forfeited upon breach, placing assets in escrow, or conducting transactions through platforms that impose rules on both sides rather than transacting privately. 
\ao{
We study this familiar commitment logic in the context of AI adoption. A user can choose an AI whose rules restrict the actions it may take on her behalf while interacting with a counterparty who has made no comparable commitment and need not enter the same contractual or institutional arrangement. The commitment is therefore attached to the intermediary the user chooses rather than to a common institution governing both parties. While this logic applies more generally to any observable commitment device that restricts its user's actions while leaving counterparties unconstrained, AI agents are the setting in which we expect such devices to become common.}

This form of commitment is especially valuable when behavior depends on beliefs about others' underlying preferences. We study this idea in a standard Prisoner's Dilemma, a simple framework that captures many applications involving exchange, information sharing, joint production, and other situations in which mutual cooperation is valuable but each side has an incentive to take advantage of a cooperative counterpart. Individuals differ in their preferences: some are selfish, while others are conditional cooperators who are willing to cooperate when they expect the other side to do the same. Because these preferences are private, even two conditional cooperators may fail to recognize one another and instead behave defensively. Cooperative preferences can therefore exist in the population without producing cooperative behavior.

A moral AI can change this informational problem. 
\ao{
We call an AI ``moral'' when it excludes defection from its feasible actions, in contrast to an unconstrained AI that leaves the user with discretion.} As a benchmark, we assume that the AI chosen by each user is observable to the other side. \jb{This assumption is natural when interactions themselves are mediated by identifiable AI agents.}
Firms can publicly disclose the systems under which their agents operate, while personal AI agents can identify themselves when communicating with other agents. Emerging standards already move in this direction. \jb{For example, the Agent2Agent (A2A) Protocol requires an agent to make available an ``Agent Card'' describing information such as its identity, provider, version, capabilities, and security requirements.\footnote{\href{https://a2a-protocol.org/dev/specification/}{https://a2a-protocol.org/dev/specification/}.}}
Such disclosure does not yet certify an agent's behavioral constraints, but it illustrates how information about the agent acting on a user's behalf can become observable to counterparties. We therefore begin with full observability to isolate the commitment channel and later relax this assumption.

Once a moral AI is observable, a user no longer needs to convince the other side that she personally intends to cooperate. 
\ao{
The technological restriction makes cooperation credible without requiring direct observation of her preferences.} In this sense, AI can substitute for information about human preferences and sustain cooperation in situations where private preferences alone cannot.

\begin{figure}[!htbp]
\centering

\begin{tikzpicture}[
    >=Latex,
    node distance=7mm,
    box/.style={
        draw=black!55,
        line width=0.6pt,
        rounded corners=2pt,
        minimum width=5.0cm,
        minimum height=0.78cm,
        align=center,
        font=\small,
        inner sep=5pt
    },
    benefit/.style={box,fill=blue!4},
    neutral/.style={box,fill=black!2},
    erosion/.style={box,fill=red!4},
    arrow/.style={-{Latex[length=2mm]},draw=black!65,line width=0.8pt}
]

\node[benefit] (a)
{Moral AI makes \textbf{cooperation credible}};

\node[benefit, below=of a] (b)
{Cooperation \textbf{expands}};

\node[neutral, below=of b] (c)
{Cooperative users become \textbf{easier to exploit}};

\node[erosion, below=of c] (d)
{
\ao{
Share of conditional cooperators \textbf{declines}}};

\node[erosion, below=of d] (e)
{Moral AI \textbf{loses its support base}};

\draw[arrow] (a) -- (b);
\draw[arrow] (b) -- (c);
\draw[arrow] (c) -- (d);
\draw[arrow] (d) -- (e);

\end{tikzpicture}

\caption{The self-liquidating mechanism. 
\ao{
Moral AI initially helps sustain cooperation, but the same commitment also changes population composition in a way that reduces support for adoption.}}
\label{fig:self_liquidating_intro}
\end{figure}
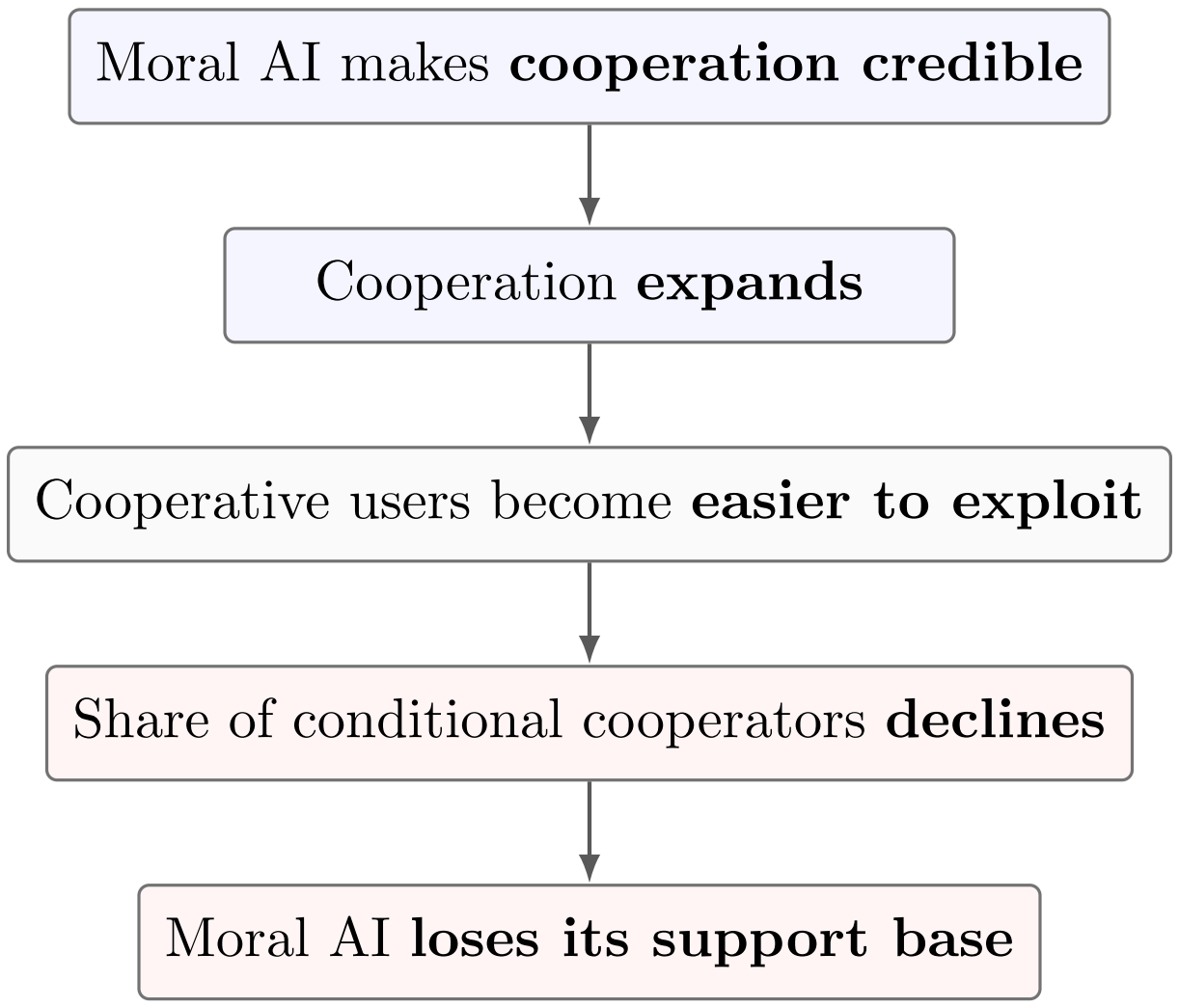

The central question of this paper is what happens when this technological solution feeds back onto human preferences. Moral AI makes cooperation possible for conditional cooperators, but the same commitment is visible to selfish individuals, who retain the option to defect and can exploit users whose AI has removed that option. The introduction of moral AI therefore creates a material advantage for selfish preferences. \jb{As individuals with higher material payoffs become more prevalent over time, the population becomes increasingly selfish.}
\ao{
This change in population composition then changes the demand for different AI technologies.} This feedback can make moral AI self-liquidating. \jb{In the baseline model, its long-run demand eventually disappears.}
\jb{The direction of this result resembles \citet{hendrycks2023natural}, but the logic is almost the reverse.}
Moral AI does not lose because constraint is inherently uncompetitive; constraint is precisely what makes it valuable in the first place. It loses because that value changes the human population on which its own demand depends. Figure \ref{fig:self_liquidating_intro} summarizes the central logic of the model.

We consider \ao{three} robustness extensions that clarify which features of the technology drive the result. First, we allow observability of AI choice to be imperfect. Each individual independently observes the opponent's AI only with some probability. \jb{This captures settings in which the AI used by a counterparty is not directly disclosed but can sometimes be inferred from its outputs: recent studies find persistent model-specific stylistic fingerprints in the outputs of different AI systems, although such inference is necessarily imperfect \citep{mcgovern2025fingerprints,lu2026stylometrics}.}
Imperfect observability weakens commitment because a constraint that is not seen cannot fully affect the opponent's behavior. 
\ao{
We show that imperfect observability weakens the benchmark effect: adoption disappears at a higher share of conditional cooperators, so selection stops at a weakly higher limiting share.}

Second, we replace perfect commitment with stochastic commitment. The moral AI normally forces cooperation but occasionally leaves the user with discretion, reflecting the possibility that its safeguards can be circumvented, overridden, or simply fail to bind in some situations. This imperfection has the opposite effect from imperfect observability. 
\ao{
As long as bypasses are rare enough that observing moral AI still induces conditional cooperators with discretion to cooperate, a small probability of bypass can make moral AI more attractive.} 
\ao{
The user continues to receive the commitment benefit because conditional-cooperator opponents respond cooperatively to the observed constraint, but sometimes regains the option to defect.} A slightly leaky constraint can therefore support commitment at an even lower population share of conditional cooperators than a perfectly binding one. 
\ao{
As a result, moral AI can remain in use at lower shares of conditional cooperators, allowing selection against them to continue further.}

\ao{Third, we relax the assumption that individuals must interact after being matched. When individuals can refuse to interact after observing their opponent's AI choice, moral AI can sustain cooperation without exposing its users to exploitation. In particular, an equilibrium exists in which both preference types adopt moral AI and refuse to interact with unconstrained AI users. Both types then receive the same material payoff, eliminating selection against conditional cooperators. However, refusal does not eliminate this selection pressure in every equilibrium. Other equilibria continue to favor selfish types, sometimes even at population shares where moral AI is not adopted in the baseline model. Thus, the evolutionary consequences of moral AI depend not only on its ability to sustain cooperation, but also on whether its users can avoid exploitation.}

Taken together, the results point to a broader question about the relationship between technology and morality. Technologies designed to support better behavior are usually judged by the behavior they produce: whether they reduce harm, sustain cooperation, or make good conduct easier. But a technology that changes the incentives surrounding moral action may also reshape the conditions under which moral dispositions are sustained, rewarded, or eroded. \jb{The long-run consequence of moral AI may thus depend not only on whether it makes people behave better, but on what kind of people remain afterward.}

\subsection*{Literature Review}
\ao{
Our model builds on the indirect evolutionary approach to preference evolution.} \citet{GuthYaari1992} and \citet{Guth1995} introduce the approach, which provides a useful theoretical framework for understanding this process. Individuals are endowed with preferences that need not coincide with material payoffs; these preferences shape their strategic behavior, while material payoffs determine evolutionary success. Preferences that induce more successful behavior therefore become more prevalent over time. Recent surveys by \citet{AlgerWeibull2019ARE} and \citet{Alger2022} review this literature and its applications to the evolution of social and other-regarding preferences. 
\ao{
A common conclusion in this literature is that, under random matching and unobservable preference types, material-payoff maximization is weakly favored \citep{GuthKliemt1998RS,Dekeletal2007RESTUD}.}

One way for non-selfish preferences to become evolutionarily valuable is for them to generate observable signals that affect how others respond. In \citet{Frank1987}, for example, a conscience acts as a commitment device, while imperfect signals of trustworthiness allow others to infer who is likely to honor that commitment. See also \cite{Akdeniz2021} on the role of commitment in the evolution of preferences. Frank also recognizes the resulting problem of mimicry: selfish individuals may imitate the signals associated with trustworthy types. \citet{RobsonandSamuelson2011} sharpen this concern, arguing that once signals themselves evolve, successful signals invite mimics, which in turn undermine their informativeness and may generate an ongoing cycle of new signals and new mimics. \citet{Hopkins2014AEJMICRO} disciplines this problem by making altruistic behavior itself a costly signal that generates better matching opportunities, while 
\ao{
\citet{heller2019coevolution} allow preferences and the ability to deceive others about those preferences to coevolve.}

A related way to sustain non-selfish preferences is to change who interacts with whom. Assortative or endogenous matching allows cooperative types to interact disproportionately with one another and thereby enjoy the benefits of their behavior \citep{AlgerAndWeibull2013ECMA,wangwu2026}. Screening institutions can achieve a similar form of separation by inducing different preference types to self-select into different groups or arrangements even when preferences themselves are unobservable \citep{wuxiong2026,HoldahlWu2026}.

Our mechanism is different. \cl{In the baseline model, matching remains random and human preferences remain hidden; moral AI neither sorts individuals nor screens them into separate groups.} 
Its observable choice may convey information about the user, but it is also a commitment device in its own right. In the benchmark model, a selfish individual can mimic a conditional cooperator's choice of moral AI only by accepting the same restriction on subsequent behavior. Moral AI therefore ties the observable signal to the behavior it promises, eliminating the possibility of appearing committed while retaining the option to defect. At the same time, because it does not shield conditional cooperators from selfish counterparts, this commitment exposes them to the selection pressure at the center of our analysis.

\ao{\citet{ElySzentes2024} also study evolutionary forces acting on AI.} They model AI embedded in machines that reproduce copies of themselves, so self-promoting AI can spread at the expense of efficient production, especially when AI is imperfectly transparent. Our mechanism is different: AI does not reproduce on its own, but changes in prevalence through human adoption as human preferences evolve. Observability also plays a different role, creating commitment value for moral AI rather than disciplining misaligned AI.

A separate literature studies how delegation and machine-mediated commitment can sustain cooperation in one-shot interactions. \citet{FershtmanJuddKalai1991} show that when principals delegate decisions to agents under publicly observable contracts, the contracts can serve as strategic commitments and support cooperative outcomes that would not arise in the underlying noncooperative game. \citet{Tennenholtz2004} develops a related idea in a computational setting: players submit programs that can inspect the programs submitted by others and condition their actions accordingly. \jb{Such programs can sustain mutual cooperation in the one-shot Prisoner's Dilemma and, more generally, generate the set of feasible and individually rational payoffs.}
\citet{CooperOesterheldConitzer2025} develop more robust simulation-based versions of program equilibrium, characterizing when similar cooperation can be supported without relying on exact syntactic comparison of programs. These papers establish that delegation and software-embedded commitment can themselves create cooperation. Our focus is instead on what happens when heterogeneous humans endogenously choose between a constrained and an unconstrained AI, and when the resulting strategic behavior feeds back into the distribution of human preferences and hence into future demand for the commitment technology.

Our analysis is also related to the emerging behavioral literature on AI delegation and morality. \citet{KobisBonnefonRahwan2021} emphasize that AI can influence ethical behavior in several roles, particularly as a partner or delegate that enables individuals to obtain unethical benefits while distancing themselves from the underlying behavior. \citet{KobisEtAl2025} provide experimental evidence that the form of machine delegation matters: people request more dishonest behavior when delegation allows them to specify high-level goals or training examples rather than directly instructing each action, while machine agents are also substantially more willing than human delegates to carry out fully unethical instructions. Our mechanism concerns a different channel. Human preferences are fixed within an interaction, and we do not model delegation as changing an individual's psychological cost of moral or immoral behavior. Instead, moral AI changes the feasible actions and therefore the strategic responses and material payoffs generated by those preferences. The resulting change in population composition is therefore an evolutionary effect rather than a within-person erosion of moral preferences.

\citet{ZimmaroEtAl2024} provide a particularly close connection to the evolutionary part of our analysis. They use evolutionary game theory to study how Samaritan and Discriminatory AI agents affect the emergence of cooperation among human-like agents in a one-shot Prisoner's Dilemma. Samaritan AIs cooperate unconditionally, whereas Discriminatory AIs condition their behavior on whether the counterpart is cooperative, and the two designs can have different effects on the evolution of cooperation. In our model, by contrast, AI is not an additional population of strategic opponents. Humans choose AI agents to act on their behalf in human--human interactions, and the prevalence of each AI technology is itself determined by human adoption. What evolves is the distribution of underlying human preferences. This distinction allows a technology that initially increases cooperation to reduce the prevalence of cooperative preferences and ultimately undermine its own demand.

\section{The Static Model}

\subsection{Preferences and AI technologies}

A unit mass of individuals is randomly matched to play a one-shot symmetric game with material payoffs of a Prisoner's Dilemma. Since the game is symmetric, we only present the material payoffs for the row player in Table \ref{material table}.

\begin{table}[ht]
\begin{center}
\caption{\label{material table} A symmetric 2 $\times$ 2 Prisoner's Dilemma}
\begin{tabular}
[c]{rcc}
    & $\text{C}$ & $\text{D}$ \\
   \cline{2-3}
    $\text{C}$ & \multicolumn{1}{|c|}{$R$} &  \multicolumn{1}{|c|}{$S$} \\
       \cline{2-3}
    $\text{D}$ & \multicolumn{1}{|c|}{$T$} &  \multicolumn{1}{|c|}{$P$} \\
       \cline{2-3}
       
\end{tabular}
\end{center}
\end{table}

The payoffs satisfy $T>R>P>S$. Action \(C\) denotes cooperation, while action \(D\) denotes defection.

There are two human preference types.  A fraction $x \in [0, 1]$
are conditional cooperators, denoted type $\theta$, and the remaining fraction $1-x$ are selfish, denoted type $\tau$. Throughout the paper, we assume that human preference types are private but individuals know the distribution of them.

A selfish type evaluates outcomes according to material payoffs. A conditional cooperator obtains an additional utility benefit $b>0$ from mutual cooperation. Hence, if a conditional cooperator is the row player, her utilities from different outcomes of the game are given in Table \ref{utility table}.
\begin{table}[ht]
\begin{center}
\caption{\label{utility table} A conditional cooperator's utilities in Prisoner's Dilemma}
\begin{tabular}
[c]{rcc}
    & $\text{C}$ & $\text{D}$ \\
   \cline{2-3}
    $\text{C}$ & \multicolumn{1}{|c|}{$R+b$} &  \multicolumn{1}{|c|}{$S$} \\
       \cline{2-3}
    $\text{D}$ & \multicolumn{1}{|c|}{$T$} &  \multicolumn{1}{|c|}{$P$} \\
       \cline{2-3}
       
\end{tabular}
\end{center}
\end{table}

We assume $R+b>T$. 
\ao{
A $\theta$-type agent's best response is $C$ if and only if she believes that her opponent chooses $C$ with probability at least $\kappa=\frac{P-S}{R+b-T+P-S}$. It is the unique best response when that probability exceeds $\kappa$; at equality, $D$ is also optimal.} 
\ao{
This makes them conditional cooperators, who cooperate when they believe their opponents are sufficiently likely to cooperate. The experimental evidence in \citet{FischbacherGaechterFehr2001} motivates this type, though it does not pin down the utility specification used here.}

In the future environment we have in mind, individuals use AI agents to act on their behalf in strategic interactions. The actions \(C\) and \(D\) should be understood as potentially complicated courses of action rather than simple button presses. An individual may decide which action she wants to take and instruct the AI to carry it out. Alternatively, she may communicate her preferences and beliefs to the AI and let it choose the action she herself would prefer under the circumstances. In either case, the underlying preferences remain those of the human; the AI is the agent through which those preferences are translated into behavior. Such an arrangement seems increasingly plausible as AI agents become capable of carrying out  complex tasks on behalf of their users.

There are two AI technologies. The moral AI permits only cooperation $M:\{C\}$, whereas the unconstrained AI permits both actions $A:\{C,D\}$. The term ``moral AI'' refers only to this restriction on the feasible action set; the AI itself need not possess preferences. In the baseline model, AI choices are perfectly observable.

\ao{
The timing is as follows. After random matching, individuals simultaneously choose an AI. They observe the opponent's choice and then simultaneously choose feasible actions. The payoff structure and the rules of each AI are common knowledge, and any randomization is independent and private.} We assume that AI choice is costless. This allows us to focus on the strategic effect of the constraint rather than differences in adoption costs.

Let $\Pr(M\mid \theta)$ and $\Pr(M\mid \tau)$ denote the adoption rate of moral AI by the two types, respectively. Let $y=x \Pr(M\mid \theta)+(1-x)\Pr(M\mid \tau)$
denote the population share using $M$.

\begin{remark}
Before studying AI-mediated strategic interactions, it is useful to consider the benchmark environment in which no AI is available and individuals therefore face no technological constraint on their actions. Type $\tau$ always defects because $D$ is its dominant action. An all-defect Bayesian Nash equilibrium therefore exists for every $x$. A second Bayesian Nash equilibrium, in which type $\theta$ cooperates and type $\tau$ defects, exists if and only if $x \ge \kappa$. Thus, when $x<\kappa$, the unique equilibrium without AI is all defection, and both preference types receive the same material payoff $P$. In this sense, cooperative preferences are present but remain dormant: they do not translate into cooperative behavior.
\end{remark}

\subsection{AI-Mediated Strategic Interactions}
In this section, we characterize the equilibria when individuals use AI agents to carry out their choices. 
\ao{
We consider \jb{type-homogeneous} Perfect Bayesian equilibria, \cl{in which all individuals of the same preference type use the same strategy.} A strategy for each human type $i\in\{\theta,\tau\}$ specifies the probability $\Pr(M\mid i)$ of choosing moral AI and, conditional on choosing $A$, a probability of choosing $C$ after each observed AI choice of the opponent. Lemma~\ref{lem:nomix} below shows that in the baseline model no equilibrium involves an interior continuation mixture on the equilibrium path.}
An individual who chooses $M$ has no subsequent action choice because $M$ permits only $C$.

After observing the opponent's AI choice, individuals update their beliefs about the opponent's preference type. By Bayes' rule, whenever $A$ is chosen with positive probability, the posterior probability that an $A$-user is type $\theta$ is
\[
\mu_A=
\frac{x(1-\Pr(M\mid\theta))}
{x(1-\Pr(M\mid\theta))+(1-x)(1-\Pr(M\mid\tau))}.
\]

 An equilibrium requires that continuation actions are optimal given these beliefs and the constraints imposed by the chosen AI, and that each type's AI choice is optimal given the resulting continuation play.

Because $M$ forces its user to choose $C$, beliefs about the preference type of an $M$-user do not affect continuation behavior in the baseline model. A $\theta$-type individual who uses $A$ and faces an $M$-user knows that the opponent will choose $C$, and therefore chooses $C$ because $R+b>T$. A $\tau$-type individual who uses $A$ always chooses $D$.

\ao{
\begin{lemma}\label{lem:nomix}
Let $0<x<1$, and allow continuation strategies to be mixed. In every \jb{type-homogeneous} Perfect Bayesian equilibrium, $\Pr(M\mid\tau)=0$, and a $\theta$-type $A$-user who meets an $A$-user on the equilibrium path chooses $C$ with probability zero or one.
\end{lemma}
\noindent\textit{Proof:} See Appendix. 
}

Continuation play can therefore take one of two forms. In the \emph{defensive continuation equilibrium}, a $\theta$-type $A$-user chooses $D$ against another $A$-user. This continuation equilibrium always exists, because if all $\theta$-type $A$-users choose $D$, an $A$-user expects no cooperation from another $A$-user. In the \emph{cooperative continuation equilibrium}, a $\theta$-type $A$-user chooses $C$ against another $A$-user. Since $\tau$-type $A$-users choose $D$, this continuation equilibrium exists if and only if $\mu_A\geq\kappa$.

\begin{proposition} \label{Prop: PBE}
\ao{
For a fixed population share $0<x<1$, all \jb{type-homogeneous} Perfect Bayesian equilibria satisfy $\Pr(M\mid\tau)=0$ and are characterized as follows.}

\begin{enumerate}
\item[(1)] \textbf{Defensive continuation equilibrium.} A $\theta$-type individual using $A$ chooses $C$ against $M$ and $D$ against $A$, while a $\tau$-type individual using $A$ chooses $D$ against either AI. Moreover, 
\[
\Pr(M\mid\theta)
=
\begin{cases}
0,
& \text{if }
x\leq \dfrac{P-S}{R+b-S},
\\[3mm]
1-\dfrac{(1-x)(P-S)}
{x(R+b-P)},
&
\text{otherwise}.
\end{cases}
\]

\item[(2)] \textbf{Cooperative continuation equilibrium.} A $\theta$-type individual using $A$ chooses $C$ against either AI, while a $\tau$-type individual using $A$ chooses $D$ against either AI. Such an equilibrium exists if and only if $x\geq\kappa$.
In every such equilibrium, 
$0\leq \Pr(M\mid\theta)
\leq
1-\frac{\kappa(1-x)}
{(1-\kappa)x}.$
\end{enumerate}
\end{proposition}

\noindent
\textit{Proof:} See Appendix. 
\\

\ao{
 Note that at the endpoints only one type is present.
At $x=0$ every individual is selfish and uses $A$. At $x=1$ several adoption patterns support cooperation. We will see that both endpoints are stationary under the selection dynamics of Section 3.}

Proposition \ref{Prop: PBE} shows that moral AI can make cooperative preferences behaviorally relevant even when they would otherwise remain dormant. Without AI, when the share of conditional cooperators is below \(\kappa\), they defect because they cannot be sufficiently confident that their opponent will cooperate. 
\ao{
\cl{Note that $\frac{P-S}{R+b-S}<\kappa$.} With observable moral AI, in the activation region $\frac{P-S}{R+b-S}<x<\kappa$, the unique equilibrium has some conditional cooperators choose $M$, which credibly commits them to cooperation and induces other conditional cooperators to cooperate in response. Below $\frac{P-S}{R+b-S}$ nobody adopts $M$, and for $x\geq\kappa$ some cooperative-continuation equilibria also have zero adoption.} As a result, cooperation can arise for population shares strictly below the threshold ($\kappa$) required without AI. 
\ao{
The proposition therefore identifies a role for observable AI commitment in activating otherwise dormant cooperation: by restricting the actions available to its user, AI can substitute for otherwise unavailable information about intentions and support cooperation that hidden human preferences alone cannot sustain.}

\section{Evolution of preferences}

We now allow the population share of conditional cooperators to evolve. At each \(x\), individuals play a \jb{type-homogeneous} Perfect Bayesian equilibrium according to their preferences on a faster time scale, while preferences evolve more slowly according to the material payoffs earned by each preference type.
This follows the standard indirect evolutionary approach, in which preferences shape behavior while material payoffs determine evolutionary success. 

For any $\sigma\in\mathcal E(x)$, 
\ao{
where $\mathcal E(x)$ denotes the set of \jb{type-homogeneous} Perfect Bayesian equilibria at $x$,
} let $F^\theta(\sigma,x)$ and $F^\tau(\sigma,x)$ denote the material fitness of the two preference types, and define the fitness differential
\begin{equation*}
\Delta(\sigma,x)\equiv F^\theta(\sigma,x)-F^\tau(\sigma,x).
\end{equation*}

\ao{
\begin{definition}\label{def:payoff-monotone}
A \emph{selection dynamic} is an autonomous law $\dot x=V(x)$ on $[0,1]$, with $V$ continuous on $[0,1]$ and $V(0)=V(1)=0$. It is \emph{payoff monotone} if, for every $x\in(0,1)$,\par
(i) if $\Delta(\sigma,x)\leq0$ for every $\sigma\in\mathcal E(x)$, then $V(x)\leq0$, with $V(x)<0$ if every such inequality is strict;\par
(ii) if $\Delta(\sigma,x)\geq0$ for every $\sigma\in\mathcal E(x)$, then $V(x)\geq0$, with $V(x)>0$ if every such inequality is strict.\par
In particular, $V(x)=0$ when every equilibrium gap is zero. If equilibrium gaps have opposing signs, the definition leaves the sign of $V(x)$ unrestricted.\par
\end{definition}

The definition deliberately does not impose an equilibrium-selection rule. When different equilibria strictly favor different types, the direction of evolution is indeterminate.
When the equilibrium is unique, the definition reduces to the usual payoff-monotonicity condition, $\operatorname{sign}V(x)=\operatorname{sign}\Delta(\sigma,x)$. 
}

\begin{proposition} \label{prop:evo}
$\dot x<0$ if and only if $1>x>\frac{P-S}{R+b-S}$, and $\dot x=0$ otherwise. Hence, for $x(0) \in (0, 1)$, $\lim_{t \rightarrow \infty}x(t)=\min\{x(0), \frac{P-S}{R+b-S}\}$ and $\lim_{t \rightarrow \infty}y(t)=0$.
\end{proposition}

\noindent
\textit{Proof:} See Appendix.
\\

Proposition \ref{prop:evo} reveals a self-liquidating force created by moral AI. By making cooperation credible, moral AI allows conditional cooperators to behave cooperatively in situations where they would otherwise defect. Yet this same commitment changes the material returns to preference types: conditional cooperators are more exposed to exploitation, while selfish individuals retain the flexibility to take advantage of them. Moral AI therefore creates selection pressure against the preferences that make it valuable in the first place. 
\ao{
As conditional cooperators become less common, the demand for commitment falls. Once the trajectory enters the unique defensive region below $\kappa$, adoption declines with $x$ and converges to zero, and the technology loses the strategic role that initially sustained its use.} The key economic implication is that a technology designed to support cooperative behavior can endogenously undermine the behavioral type on which its own value depends. Figure \ref{fig:selfliquidating} provides a graphic illustration. 

\ao{
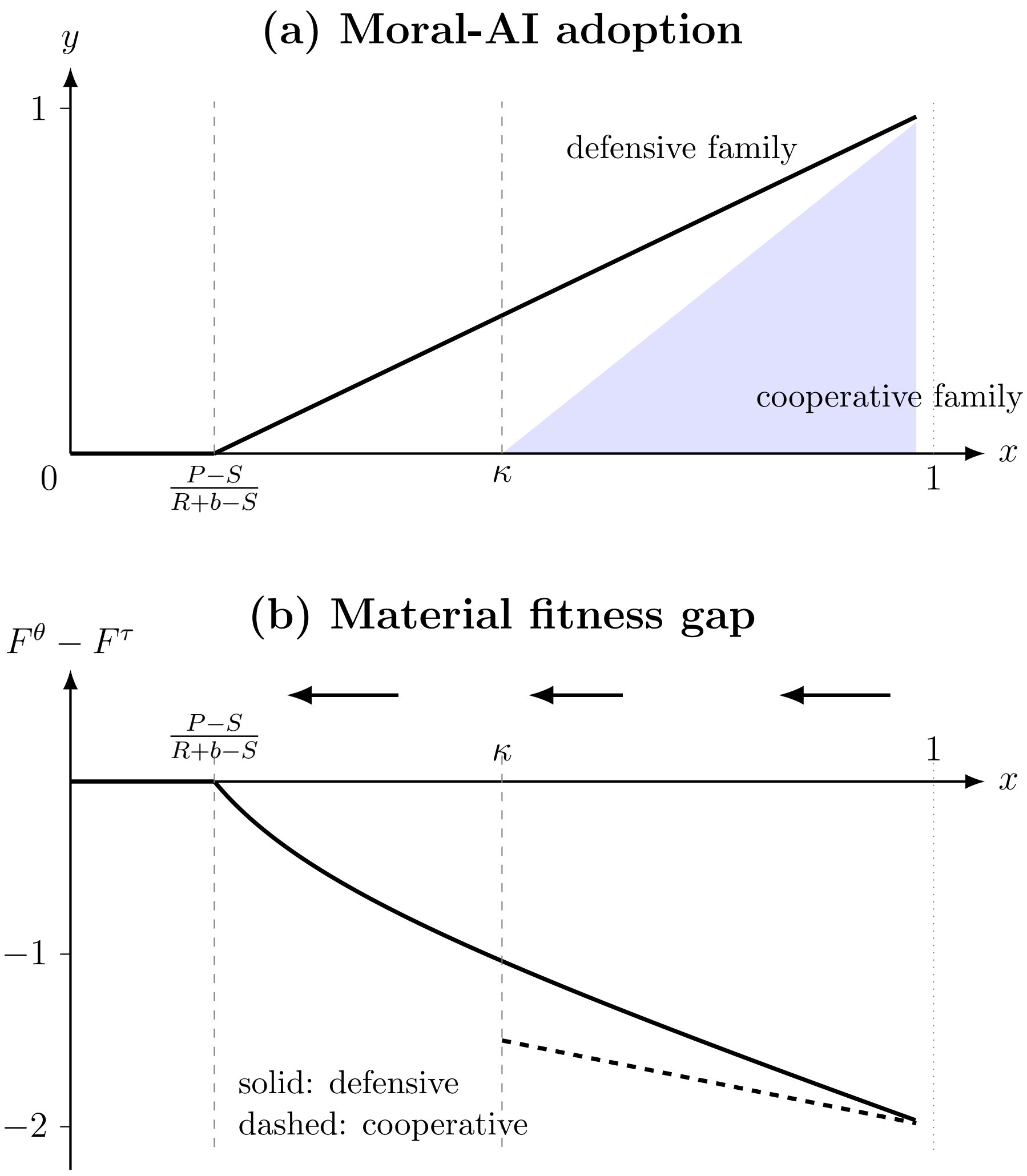
\begin{figure}[!htbp]
\centering
\begin{tikzpicture}[x=10cm,y=4cm,>=Latex,every node/.style={font=\normalsize}]
\begin{scope}[shift={(0,0.95)}]
\node[font=\large\bfseries] at (0.5,1.22) {(a) Moral-AI adoption};
\fill[blue!12] (0.5,0) -- plot[domain=0.5:0.98,samples=2] (\x,{2*\x-1}) -- (0.98,0) -- cycle;
\draw[->,thick] (0,0) -- (1.06,0) node[right] {$x$};
\draw[->,thick] (0,0) -- (0,1.12) node[above] {$y$};
\draw[line width=1.4pt] (0,0) -- (1/6,0);
\draw[line width=1.4pt] plot[domain=1/6:0.98,samples=2] (\x,{(6*\x-1)/5});
\draw[dashed,gray] (1/6,0) -- (1/6,1.02);
\draw[dashed,gray] (0.5,0) -- (0.5,1.02);
\draw[dotted,gray] (1,0) -- (1,1.02);
\node[below] at (1/6,0) {$\frac{P-S}{R+b-S}$};
\node[below] at (0.5,0) {$\kappa$};
\node[below] at (1,0) {$1$};
\node[below left] at (0,0) {$0$};
\draw (0,1) -- (-0.012,1) node[left] {$1$};
\node[anchor=west,font=\small] at (0.56,0.88) {defensive family};
\node[anchor=west,font=\small] at (0.78,0.16) {cooperative family};
\end{scope}
\begin{scope}[yscale=0.5]
\node[font=\large\bfseries] at (0.5,0.95) {(b) Material fitness gap};
\draw[->,thick] (0,0) -- (1.06,0) node[right] {$x$};
\draw[->,thick] (0,-2.25) -- (0,0.65) node[above] {$F^\theta-F^\tau$};
\draw[line width=1.4pt] (0,0) -- (1/6,0);
\draw[line width=1.4pt,smooth] plot[domain=1/6:0.98,samples=80] (\x,{-(6*\x-1)*(3+7*\x)/(25*\x)});
\draw[line width=1.4pt,dashed] plot[domain=0.5:0.98,samples=2] (\x,{-1-\x});
\draw[dashed,gray] (1/6,0.15) -- (1/6,-2.15);
\draw[dashed,gray] (0.5,0.15) -- (0.5,-2.15);
\draw[dotted,gray] (1,0.15) -- (1,-2.15);
\node[above] at (1/6,0.05) {$\frac{P-S}{R+b-S}$};
\node[above] at (0.5,0.05) {$\kappa$};
\node[above] at (1,0.05) {$1$};
\draw (0,-1) -- (-0.012,-1) node[left] {$-1$};
\draw (0,-2) -- (-0.012,-2) node[left] {$-2$};
\draw[->,very thick] (0.95,0.5) -- (0.82,0.5);
\draw[->,very thick] (0.64,0.5) -- (0.53,0.5);
\draw[->,very thick] (0.38,0.5) -- (0.25,0.5);
\node[anchor=north west,font=\small,align=left] at (0.18,-1.6) {solid: defensive\\dashed: cooperative};
\end{scope}
\end{tikzpicture}
\caption{\ao{Equilibrium adoption and material fitness under observable commitment. In the defensive continuation equilibrium, moral-AI adoption is positive for $x>\frac{P-S}{R+b-S}$. For $x\geq\kappa$, cooperative-continuation equilibria additionally permit every adoption rate in the shaded set, including zero. At each interior state $x>\frac{P-S}{R+b-S}$, every equilibrium favors selfish types in material fitness. Trajectories starting in $(\frac{P-S}{R+b-S},1)$ converge to $\frac{P-S}{R+b-S}$, with adoption converging to zero. Drawn for $T=5$, $R=3$, $P=1$, $S=0$ and $b=3$, so $\frac{P-S}{R+b-S}=1/6$ and $\kappa=1/2$. These parameter values are illustrative.}}
\label{fig:selfliquidating}
\end{figure}
}


\section{Robustness}

\subsection{Imperfect observability}

We first maintain perfect enforcement of moral AI but relax the assumption that its use is always observed. In practice, a counterparty may know that an AI agent is acting on someone's behalf without being able to verify exactly which system is being used or what constraints govern it. Even as technologies for disclosure and verification make AI use increasingly observable, such observability is likely to remain imperfect.

Assume that each player independently observes the opponent's AI with probability $\rho\in(0,1]$. A player knows whether she observes the opponent's AI but does not know whether the opponent observes hers. We have the following result.

\begin{proposition} \label{prop:obs}
Suppose $\rho
>
\frac{R+b-T+P-S}{R+b-S}$. Then, $\dot x<0$ if and only if $1>x>\frac{P-S}{\rho(R+b-S)}$, and $\dot x=0$ otherwise. Hence, for $x(0) \in (0, 1)$, $\lim_{t \rightarrow \infty}x(t)=\min\{x(0), \frac{P-S}{\rho(R+b-S)}\}$ and $\lim_{t \rightarrow \infty}y(t)=0$. 
\end{proposition}

\noindent \textit{Proof:} See Appendix.
\\

For sufficiently high $\rho$, Proposition \ref{prop:obs} implies that imperfect observability weakens the baseline mechanism (\(\rho<1\) implies that $\frac{P-S}{\rho(R+b-S)}
>
\frac{P-S}{R+b-S}$). 
\ao{
The rationale is that the response benefit of commitment requires that the opponent observe it.} A conditional cooperator who adopts moral AI is still constrained to cooperate when her AI choice goes unobserved, but in those encounters the constraint cannot induce additional cooperation from the opponent. Imperfect observability therefore reduces the benefit of commitment without eliminating its cost. 
\ao{
As the population share of conditional cooperators falls, moral AI consequently ceases to be attractive at a higher $x$ than under perfect observability, and selection against conditional cooperators stops at a weakly higher share.} Thus greater observability has two reinforcing effects: it makes moral AI more valuable and allows it to sustain cooperation for a wider range of population states. But precisely for this reason, it also extends the range over which moral AI selects against conditional cooperators. 
\ao{
Imperfect observability dampens the self-liquidating force and leaves a weakly larger limiting share of conditional cooperators from a common initial population. The two limits are equal when the initial share lies below both cutoffs.}

\ao{
For lower observability, additional continuation equilibria may alter the adoption region and the states at which selection stops. The imitation argument nevertheless implies that selfish types weakly outperform conditional cooperators at every equilibrium. We leave the full equilibrium characterization in this parameter region to future work.}

\subsection{Stochastic commitment}

We next restore perfect observability of AI choice and consider imperfect enforcement. In practice, even an AI designed to enforce cooperation may occasionally fail, be overridden, or be circumvented: with probability $1-\varepsilon$, an $M$-user is forced to cooperate, while with probability $\varepsilon$, the user remains free to choose either action. The opponent observes that $M$ is being used and knows $\varepsilon$, but does not observe whether the restriction has been bypassed. 
\ao{
Bypass events are independent across players and independent of preference types and AI choices. Each $M$-user privately observes whether her own constraint binds before choosing an action when discretion is available. Opponents observe the $M$ label but not the bypass realization.}

\ao{
We restrict attention to sufficiently small bypass probabilities:}
\[
\varepsilon
<\bar{\varepsilon}\equiv
\min\left\{
1-\kappa,\,
\frac{R+b-P}
{(R+b-P)+(R+b-T+P-S)},\,
\frac{T-R}
{(T-P)+(T-R)}
\right\}.
\]

The first condition ensures that observing $M$ still induces a conditional opponent to cooperate even if every user who obtains a bypass defects. 
\ao{
The second makes the adoption gain strictly decreasing in $\Pr(M\mid\theta)$ in the region $x<\kappa$.} The third ensures that a selfish type does not find moral AI attractive even when all conditional cooperators adopt it.

\begin{proposition} \label{prop: stochastic}
Suppose $\varepsilon
<\bar{\varepsilon}$ holds. Then, $\dot x<0$ if and only if $1>x>\frac{
(1-\varepsilon)(P-S)
}{
(1-\varepsilon)(R+b-S)
+\varepsilon(T-P)
}$, and $\dot x=0$ otherwise. Hence, for $x(0) \in (0, 1)$,  $\lim_{t \rightarrow \infty}x(t)=\min\{x(0), \frac{
(1-\varepsilon)(P-S)
}{
(1-\varepsilon)(R+b-S)
+\varepsilon(T-P)}\}$ and $\lim_{t \rightarrow \infty}y(t)=0$. 
\end{proposition}

\noindent
\textit{Proof:} See Appendix.
\\

Proposition \ref{prop: stochastic} implies that a small stochastic failure has the opposite comparative effect from imperfect observability. Imperfect observability weakens commitment because the other side sometimes fails to see it. A small bypass leaves the commitment visible and sufficiently credible, while occasionally restoring discretion to the user. 
\ao{
Consequently,
\[
\frac{(1-\varepsilon)(P-S)}{(1-\varepsilon)(R+b-S)+\varepsilon(T-P)}
\;\leq\;
\frac{P-S}{R+b-S}
\;\leq\;
\frac{P-S}{\rho(R+b-S)}
\]
for $0\leq\varepsilon<\bar{\varepsilon}$ and $\frac{R+b-T+P-S}{R+b-S}<\rho\leq1$. The first inequality is strict when $\varepsilon>0$, and the second is strict when $\rho<1$.} 
\ao{
Imperfect observability causes moral-AI adoption to disappear at a higher population share; a small stochastic bypass instead allows it to remain attractive after conditional cooperators have become even less common.}

\ao{Note that under the restriction of Proposition~\ref{prop: stochastic}, $0\leq\varepsilon<\bar{\varepsilon}$, the adoption cutoff $\frac{
(1-\varepsilon)(P-S)
}{
(1-\varepsilon)(R+b-S)
+\varepsilon(T-P)
}$ decreases strictly with $\varepsilon$.}

We do not consider large enforcement errors. Our purpose is to examine whether the baseline mechanism survives small departures from perfect commitment. As $\varepsilon$ becomes large, moral AI provides progressively weaker commitment, and additional continuation behavior may arise. In the extreme case $\varepsilon=1$, $M$ imposes no restriction at all and becomes strategically equivalent to $A$, so the commitment channel disappears entirely.

We also leave the interaction between imperfect observability and imperfect enforcement for future work. 
\ao{
The two frictions depart from perfect observable commitment in different ways and have opposite effects on the adoption cutoff.} 
Characterizing this interaction would require jointly studying the visibility and reliability of commitment, which we leave outside the scope of the present analysis.




\ao{

\subsection{A right to refuse}

The baseline model assumes that randomly matched individuals must interact, even when their opponent's AI choice exposes them to exploitation. We now allow individuals to decline interaction. After choosing an AI and observing the opponent's choice, both individuals simultaneously decide whether to accept the match. The Prisoner's Dilemma is played only if both accept; otherwise, each receives material payoff $o<R$ and no moral benefit. Matching remains random, and there is no rematching. If $o<S$, declining is never profitable and the baseline model is recovered. We therefore consider $S<o<R$.

\begin{proposition}\label{prop:refuse}
Suppose $S<o<R$ and $0<x<1$.
\begin{enumerate}
\item[(i)] There is an equilibrium in which both types choose $M$ and decline opponents who use $A$. Both types earn material payoff $R$, so $\Delta(\sigma,x)=0$.

\cl{\item[(ii)] In every equilibrium, $F^\tau(\sigma, x)\geq F^\theta(\sigma, x)$.}

\item[(iii)] Equilibria with $\Delta(\sigma,x)<0$ also exist. If $S<o<P$, they exist for every $x>\frac{P-S}{R+b-S}$ and whenever
\[
\frac{P-o}{R+b+P-2o}\leq x<
\frac{P-o}{R+P-2o},
\]
where the lower bound is below $\frac{P-S}{R+b-S}$. If $P\leq o<R$, they exist for every
\[
x>\frac{o-S}{R+b-S}.
\]
\end{enumerate}
\end{proposition}
\noindent\textit{Proof:} See Appendix. 
\\

Allowing individuals to refuse interaction changes the evolutionary consequences of moral AI. When both types adopt $M$ and refuse to interact with $A$-users, moral AI sustains cooperation without exposing its users to exploitation. Selfish individuals also adopt $M$, since doing so allows them to participate in cooperative interactions. Both types then receive the same material payoff, eliminating selection against conditional cooperators.

However, refusal does not eliminate equilibria in which selfish individuals have a material advantage. In particular, when $S<o<P$, such equilibria can arise even below the original adoption cutoff $\frac{P-S}{R+b-S}$, where moral AI is not adopted in the baseline model. Thus, the option to refuse interaction can prevent the erosion of cooperative preferences in some equilibria while allowing it to persist in others.

Since every population state admits an equilibrium with equal material fitness, while selfish types are weakly favored in every equilibrium, Definition 1 implies $\dot{x}\leq 0$ but does not generally require strict decline. The long-run outcome therefore depends on equilibrium selection.

}

\section{Conclusion}
This paper studies how moral AI can shape both strategic behavior and the evolution of human preferences. By restricting the actions available to its user, moral AI can make cooperation credible and induce conditional cooperators to cooperate even when their preferences would otherwise remain behaviorally dormant. Yet this same commitment creates a material advantage for selfish preferences, causing the population share of conditional cooperators to fall over time. 
\ao{
As a result, for initially mixed populations, moral-AI adoption eventually converges to zero and the technology becomes self-liquidating.} 
\ao{
The mechanism survives sufficiently small departures from perfect observability and perfect enforcement, and the two frictions shift the adoption cutoff in opposite directions. Allowing individuals to refuse interaction has a different effect: moral AI can sustain cooperation without generating selection against conditional cooperators, although other equilibria continue to favor selfish types.} \jb{More broadly, the study of AI morality should ask not only how to make AI behave morally, but also how moral AI changes human behavior and preferences over time.}
\cl{As AI increasingly acts on our behalf, these feedback effects may become as important as the moral properties of the AI itself.}


\subsection*{Declaration of AI use}

 \ao{OpenAI's ChatGPT was used during the development of this study to discuss and refine ideas, improve exposition, assist with editing and \LaTeX\ formatting, prepare figures, and check the presentation of mathematical arguments. During revision, Anthropic's Claude was used to review the manuscript, propose corrections to proofs and wording, and draft inserted text. The research questions, the model and the results are the authors' own. The authors checked every proof and all AI-assisted material and take full responsibility for the final manuscript.}

\appendix
\section{Appendix}

\ao{
\subsection{Proof of Lemma \ref{lem:nomix}}
Recall that a type-$\tau$ individual using $A$ always defects, whereas a type-$\theta$ individual using $A$ cooperates against $M$, since $R+b>T$. 

We first show that $\Pr(M\mid\tau)=0$. Suppose otherwise. Choosing $M$ gives both types the same material payoff, whereas choosing $A$ gives type $\tau$ a weakly higher material payoff, since defection is materially dominant. Moreover, type $\theta$ is weakly more likely to achieve mutual cooperation by choosing $M$ rather than $A$. Thus, type $\theta$ has a weakly stronger incentive to adopt $M$ than type $\tau$. This comparison is strict whenever some individuals use $M$, \cl{since against an $M$-user, switching to $A$ gains type $\tau$ $T-R$ and gains type $\theta$ nothing.}

Since $\Pr(M\mid\tau)>0$, this implies $\Pr(M\mid\theta)=1$. If some individuals still use $A$, they must all be type $\tau$, so type $\theta$ strictly prefers $A$ to $M$, since $P>S$ against $A$-users and both technologies yield $R+b$ against $M$-users. If everyone uses $M$, type $\tau$ can profitably deviate to $A$, earning $T>R$. Both cases yield a contradiction. Hence $\Pr(M\mid\tau)=0$.

Next, suppose a type-$\theta$ $A$-user strictly mixes between $C$ and $D$ against another $A$-user. Indifference requires the opponent to cooperate with probability $\kappa$. Since type $\tau$ always defects and type $\theta$ mixes, we must have $\mu_A>\kappa$.

Against an $A$-user, the utility gain from choosing $M$ rather than $A$ is
\[
\begin{aligned}
&\mu_A(R+b)+(1-\mu_A)S\\
&\quad-\left[\kappa T+(1-\kappa)P\right]\\
&=(\mu_A-\kappa)(R+b-S)>0,
\end{aligned}
\]

where the equality follows from the definition of $\kappa$. Against an $M$-user, both technologies yield the same utility $R+b$. Therefore, type $\theta$ strictly prefers $M$, contradicting the use of $A$ on the equilibrium path. \hfill$\square$

}

\subsection{Proof of Proposition \ref{Prop: PBE}}
Consider first the defensive continuation equilibrium. 
\ao{
For a $\theta$-type individual, the payoff difference between $M$ and $A$ arises only against an $A$-user.} Against a $\theta$-type $A$-user, $M$ turns mutual defection into mutual cooperation, yielding a gain of $R+b-P$. Against a $\tau$-type $A$-user, $M$ instead exposes her to defection, yielding a loss of $P-S$. Opponents using $M$ generate no payoff difference because a $\theta$-type individual cooperates against them whether she uses $M$ or $A$. Hence her expected utility gain from $M$ over $A$ is
\[
\Delta_\theta
=
x\bigl(1-\Pr(M\mid\theta)\bigr)(R+b-P)
-
(1-x)\bigl(1-\Pr(M\mid\tau)\bigr)(P-S).
\tag{1}
\]

By Lemma \ref{lem:nomix}, $\Pr(M\mid\tau)=0$. Substituting this into (1) gives
$\Delta_\theta
=
x\bigl(1-\Pr(M\mid\theta)\bigr)(R+b-P)
-
(1-x)(P-S)$. 
If no $\theta$-type adopts $M$, $\Delta_\theta
=x(R+b-P)-(1-x)(P-S)$, 
which is nonpositive if and only if $x\leq\frac{P-S}{R+b-S}$. Thus $\Pr(M\mid\theta)=0$ for population shares below this threshold.

When $1>x>\frac{P-S}{R+b-S}$, 
a $\theta$-type strictly prefers $M$ when $\Pr(M\mid\theta)=0$, but strictly prefers $A$ when $\Pr(M\mid\theta)=1$.
Hence both technologies must be used, and indifference requires $\Delta_\theta=0$, which gives
$\Pr(M\mid\theta)
=
1-
\frac{(1-x)(P-S)}
{x(R+b-P)}$.
This establishes the defensive equilibrium.

Now consider the cooperative continuation equilibrium. A $\theta$-type individual using $A$ chooses $C$ against either technology. She therefore obtains exactly the same outcome whether she chooses $M$ or $A$ and is indifferent between them. 
By Lemma \ref{lem:nomix}, $\Pr(M\mid\tau)=0$. Hence, by Bayes' rule, the posterior probability that an $A$-user is type $\theta$ is
$\mu_A
=
\frac{x(1-\text{Pr}(M\mid\theta))}
{x(1-\text{Pr}(M\mid\theta))+(1-x)}$.

For a $\theta$-type $A$-user to cooperate against another $A$-user, sequential rationality requires $\mu_A\geq\kappa$. This implies $\Pr(M\mid\theta)
\leq
1-
\frac{\kappa(1-x)}
{(1-\kappa)x},
$
whose right hand side is nonnegative if and only if $x\geq\kappa$. Hence the cooperative continuation equilibrium exists if and only if $x\geq\kappa$, and whenever it exists any $0\leq\Pr(M\mid\theta)
\leq
1-
\frac{\kappa(1-x)}
{(1-\kappa)x}$
can be sustained.

This exhausts all \jb{type-homogeneous} Perfect Bayesian equilibria.
\hfill$\square$

\subsection{Proof of Proposition \ref{prop:evo}}
Consider first the defensive continuation equilibrium. Since $\Pr(M\mid \tau)=0$, a type-$\theta$ user of $M$ obtains material payoff $xR+(1-x)S$,
while a type-$\theta$ user of $A$ obtains $x\Pr(M\mid \theta)R+(1-x\Pr(M\mid \theta))P$. In addition, a type-$\tau$ individual using $A$ obtains $x\Pr(M\mid \theta)T+(1-x\Pr(M\mid \theta))P$.

Hence
\[
\begin{split}
\Delta(\sigma,x)
={}&
\Pr(M\mid \theta)[xR+(1-x)S]
 \\
&+(1-\Pr(M\mid \theta))[x\Pr(M\mid \theta)R+(1-x\Pr(M\mid \theta))P]\\
&- \big(x\Pr(M\mid \theta)T+(1-x\Pr(M\mid \theta))P\big).
\end{split} \tag{4} 
\]

When $x\leq\frac{P-S}{R+b-S}$, $\Pr(M\mid \theta)=0$. We have $\Delta(\sigma,x)=0$. When
$x>\frac{P-S}{R+b-S}$, $\Pr(M\mid \theta)=
1-\frac{(1-x)(P-S)}
{x(R+b-P)}$. This implies $
\Delta(\sigma,x)
=
-\Pr(M\mid \theta)\left[
\frac{(1-x)(P-S)b}{R+b-P}
+x(T-R)
\right]
<0.$

On the cooperative continuation branch, every type-$\theta$ individual cooperates and every type-$\tau$ individual defects. Thus,  $F^\theta(\sigma,x)=xR+(1-x)S$
and
$F^\tau(\sigma,x)=xT+(1-x)P$,
implying that 
$\Delta(\sigma,x)
=x(R-T)+(1-x)(S-P)<0.$

When $\frac{P-S}{R+b-S}<x<\kappa$,
only the defensive continuation equilibrium exists. Since $\Delta(\sigma,x)<0$, evolution strictly favors type $\tau$ over type $\theta$. 
\ao{
For $x\geq\kappa$, both continuation families may exist, but both give the same direction of selection.} Below the cutoff $\frac{P-S}{R+b-S}$, moral AI is not adopted, all individuals defect, and selection stops.


\ao{
It remains to establish the limits. If $x(0)\leq \frac{P-S}{R+b-S}$, then $V(x(0))=0$ and the trajectory is constant. If $x(0)\in(\frac{P-S}{R+b-S},1)$, the trajectory decreases while $x>\frac{P-S}{R+b-S}$ and cannot cross the interval $[0,\frac{P-S}{R+b-S}]$ on which $V$ vanishes, so it has a limit $\ell\geq \frac{P-S}{R+b-S}$. If $\ell>\frac{P-S}{R+b-S}$, then $V(\ell)<0$, and by continuity $V$ is bounded away from zero near $\ell$, which contradicts convergence. Hence $\ell=\frac{P-S}{R+b-S}$. Because $\frac{P-S}{R+b-S}<\kappa$, the trajectory eventually enters $(\frac{P-S}{R+b-S},\kappa)$, where the defensive equilibrium is unique and
\[
y=\frac{(R+b-S)x-(P-S)}{R+b-P},
\]
which tends to zero as $x\to \frac{P-S}{R+b-S}$. Both limits therefore hold whichever equilibrium is played at $x\geq\kappa$.}

\hfill$\square$

\subsection{Proof of Proposition \ref{prop:obs}}
First, observe that $\rho > \frac{R+b-T+P-S}{R+b-S}$ implies $ \frac{P-S}{\rho(R+b-S)}<\kappa$.

\ao{As in the proof of Lemma \ref{lem:nomix}, type $\theta$ has a weakly stronger incentive to adopt $M$ than type $\tau$. This comparison continues to hold under imperfect observability, since an unobserved AI choice does not affect the opponent's behavior, while an observed $M$ induces cooperation by type-$\theta$ $A$-users. The comparison is strict whenever $\Pr(M\mid\tau)>0$, since $\rho>0$ and type $\theta$ cooperates against an observed $M$-user while type $\tau$ defects. Thus, $\Pr(M\mid\tau)>0$ implies $\Pr(M\mid\theta)=1$. But then all $A$-users are selfish, and type $\tau$ strictly prefers $A$ to $M$, a contradiction. Hence $\Pr(M\mid\tau)=0$.}

Now consider $x<\kappa$. After observing $A$, the posterior probability that the opponent is type $\theta$ is no greater than $x<\kappa$. If the opponent's AI is not observed, the probability of cooperation is also at most $x<\kappa$. Thus a $\theta$-type $A$-user cooperates only when she observes $M$ and defects otherwise.

Against a type-$\theta$ $A$-user, choosing $M$ induces cooperation only when it is observed, yielding a gain of $\rho(R+b)+(1-\rho)S-P$. Against a type-$\tau$ $A$-user, $M$ only exposes the individual to defection, yielding a loss of $P-S$. Against a type-$\theta$ $M$-user, choosing $M$ preserves mutual cooperation, whereas choosing $A$ allows defection when $M$ is not observed. Hence, the gain from choosing $M$ rather than $A$ in this case is $R+b-[\rho(R+b)+(1-\rho)T]=(1-\rho)(R+b-T)$. This implies that the expected utility gain from choosing $M$ rather than $A$ is
\[
\begin{split}
\Delta_\theta
={}&
x\bigl(1-\Pr(M\mid\theta)\bigr)
\left[\rho(R+b)+(1-\rho)S-P\right]\\
&-(1-x)(P-S)
+x\Pr(M\mid\theta)(1-\rho)(R+b-T),\\
={}&
\rho x(R+b-S)-(P-S)\\
&+
x\Pr(M\mid\theta)
\left[
(1-\rho)(R+b-T+P-S)-\rho(R+b-P)
\right].
\end{split}
\tag{5}
\]

Given $\rho
>
\frac{R+b-T+P-S}{R+b-S}$, the coefficient on $\Pr(M\mid\theta)$ is strictly negative. At $\Pr(M\mid\theta)=0$, $\Delta_\theta=\rho x(R+b-S)-(P-S)$. Hence no type $\theta$ adopts $M$ if and only if $x\leq\frac{P-S}{\rho(R+b-S)}$.

\ao{Above this cutoff and below $\kappa$, equation (5) implies that when $\Pr(M\mid\theta)=0$, the gain from adopting $M$ is $\rho x(R+b-S)-(P-S)>0$. When $\Pr(M\mid\theta)=1$, this gain is $x(1-\rho)(R+b-T)-(1-x)(P-S)<0$, where the inequality follows from $x<\kappa$. Thus neither zero nor full adoption can be an equilibrium, implying $0<\Pr(M\mid\theta)<1$. Moreover the material-fitness comparison has the same sign as in the baseline model. Since $0<\Pr(M\mid\theta)<1$, some type-$\theta$ individuals use $A$ and cooperate when they observe $M$. Because preference types are private, type $\tau$ can imitate type $\theta$'s AI-adoption strategy while defecting whenever discretion is available, thereby obtaining a strictly higher material payoff. Hence $F^\tau(\sigma,x)>F^\theta(\sigma,x)$.}

\ao{The same argument applies to the additional equilibria that may exist for $x\geq\kappa$. If $\Pr(M\mid\theta)=0$, some type-$\theta$ individuals must cooperate, since otherwise adopting $M$ would be profitable above the cutoff. If $0<\Pr(M\mid\theta)<1$, type-$\theta$ individuals using $A$ cooperate when they observe $M$. In both cases, type $\tau$ can imitate type $\theta$'s AI choice and earn a strictly higher material payoff by defecting whenever discretion is available. Finally, if $\Pr(M\mid\theta)=1$, the material payoffs are $xR+(1-x)S$ and $xT+(1-x)P$ for the two types, respectively, so type $\tau$ again has a strict advantage.}

\cl{Thus $\dot x<0$ above the cutoff $\frac{P-S}{\rho(R+b-S)}$, while below it everyone uses $A$, everyone defects, and the two types earn the same material payoff; convergence follows as in the proof of Proposition~\ref{prop:evo}.}
\hfill$\square$

\subsection{Proof of Proposition \ref{prop: stochastic}}
First, $\varepsilon
<\bar{\varepsilon}$ implies $1-\varepsilon>\kappa$.
Hence a $\theta$-type $A$-user cooperates after observing $M$ even if every $M$-user who obtains a bypass subsequently defects.


\ao{
Second, we show that type $\tau$ does not adopt $M$. Suppose $\Pr(M\mid\tau)>0$. As in the proof of Lemma \ref{lem:nomix}, we compare the two types' incentives to choose $M$ rather than $A$. Against $A$-users, this incentive is weakly stronger for type $\theta$. If a type-$\theta$ $A$-user defects or mixes against $A$, her expected utility from $A$ is the same as that of type $\tau$, while choosing $M$ gives type $\theta$ at least as much utility as type $\tau$. If she cooperates against $A$, the difference between the two types' adoption incentives is
\[
(1-\varepsilon)[\mu_A(T-R)+(1-\mu_A)(P-S)]>0.
\]
Against an $M$-user, the difference is also strictly positive: it is at least $(1-\varepsilon)(T-R)>0$ against type $\theta$, and equals $(1-\varepsilon)[(1-\varepsilon)(T-R)+\varepsilon(P-S)]>0$ against type $\tau$. Thus, type $\theta$ has a strictly stronger incentive to adopt $M$ than type $\tau$, implying $\Pr(M\mid\theta)=1$.

But when all type-$\theta$ individuals choose $M$, type $\tau$ strictly prefers $A$. Against a type-$\theta$ $M$-user, the gain from switching to $A$ is at least
\[
(1-\varepsilon)(T-R)-\varepsilon(T-P)>0,
\]
where the inequality follows from the third restriction on $\varepsilon$. Against a type-$\tau$ $A$-user, the gain is $(1-\varepsilon)(P-S)>0$, and against a type-$\tau$ $M$-user, it is $(1-\varepsilon)[(1-\varepsilon)(T-R)+\varepsilon(P-S)]>0$. This contradicts $\Pr(M\mid\tau)>0$. Hence $\Pr(M\mid\tau)=0$.
}

Now suppose $x<\kappa$. Since $\tau$-type individuals do not adopt $M$, the posterior probability that an $A$-user is $\theta$-type is no greater than $x<\kappa$. Thus a $\theta$-type $A$-user defects against an $A$-user. By contrast, an observed $M$-user cooperates with probability at least $1-\varepsilon$. 
\ao{
Since $1-\varepsilon>\kappa$, a $\theta$-type $A$-user cooperates against $M$.} If a $\theta$-type $M$-user obtains a bypass, she follows the same continuation behavior.

Under this continuation behavior, for a $\theta$ type, against a $\theta$-type $A$-user, choosing $M$ rather than $A$ yields a gain of $(1-\varepsilon)(R+b-P)+\varepsilon(T-P)$.

Against a type $\tau$, it yields a loss of $(1-\varepsilon)(P-S)$, while against a $\theta$-type $M$-user it yields a gain of $\varepsilon(R+b-S)$. 

The expected utility gain from $M$ rather than $A$ for a $\theta$ type can therefore be written as
\[
\begin{split}
\Delta_\theta
={}&
x\bigl(1-\Pr(M\mid\theta)\bigr)
\left[(1-\varepsilon)(R+b-P)+\varepsilon(T-P)\right]\\
&-(1-x)(1-\varepsilon)(P-S)
+x\Pr(M\mid\theta)\varepsilon(R+b-S)\\
={}&
x\left[(1-\varepsilon)(R+b-P)+\varepsilon(T-P)\right]
-(1-x)(1-\varepsilon)(P-S)\\
&-
x\Pr(M\mid\theta)
\left[
(1-\varepsilon)(R+b-P)
-\varepsilon(R+b-T+P-S)
\right].
\end{split}
\tag{6}
\]
$\varepsilon<\bar{\varepsilon}$ ensures that the coefficient on $\Pr(M\mid\theta)$ is strictly negative. At $\Pr(M\mid\theta)=0$, $\Delta_\theta\leq0$ if and only if
\[
x\leq
\frac{(1-\varepsilon)(P-S)}
{(1-\varepsilon)(R+b-S)+\varepsilon(T-P)}.
\]
Hence at and below this cutoff no $\theta$-type adopts $M$.


\ao{

Above the cutoff, every equilibrium involves cooperation by type $\theta$. If $\Pr(M\mid\theta)=0$, some type-$\theta$ individuals must cooperate, since otherwise adopting $M$ would be profitable. If $0<\Pr(M\mid\theta)<1$, type-$\theta$ $A$-users cooperate against $M$. Finally, if $\Pr(M\mid\theta)=1$, then $\varepsilon>0$, since otherwise type $\theta$ would prefer $A$ against selfish $A$-users. In this case, bypassed type-$\theta$ $M$-users cooperate against $M$, since $1-\varepsilon>\kappa$. Thus, in every case, type $\theta$ cooperates with positive probability when discretion is available. Because preferences are private, type $\tau$ can imitate type $\theta$'s AI-adoption strategy while defecting whenever discretion is available, obtaining a strictly higher material payoff.

At or below the cutoff, equation (6) implies that no one adopts $M$. Since the cutoff is below $\kappa$, everyone defects and both types receive the same material payoff. The convergence result then follows as in the proof of Proposition \ref{prop:evo}.
}



\hfill$\square$

\ao{
\subsection{Proof of Proposition \ref{prop:refuse}}

(i) Consider an equilibrium in which both types choose $M$, accept $M$-users, and decline $A$-users. Since $A$ is off the equilibrium path, assign probability zero to an $A$-user being type $\theta$. An $M$-user facing an $A$-user therefore expects defection and prefers to decline, since $o>S$. Against another $M$-user, both types prefer to accept because $R>o$. When both matched individuals choose $A$, they defect if the match proceeds and accept the match if and only if $P\geq o$. A deviation to $A$ is rejected by every $M$-user and yields $o<R$. Hence neither type has an incentive to deviate, and both receive material payoff $R$.

\cl{(ii) As in the previous environments, type $\tau$ can imitate the AI choice and acceptance strategy of type $\theta$ while defecting whenever discretion is available. Since defection is materially dominant, this implies $F^\tau(\sigma,x)\geq F^\theta(\sigma,x)$ in every equilibrium.}


(iii) First, suppose $S<o<P$ and $x>\frac{P-S}{R+b-S}$. Consider the defensive continuation equilibrium of Proposition \ref{Prop: PBE}, with every match accepted. The posterior probability that an $A$-user is type $\theta$ is $\mu_A=\frac{P-S}{R+b-S}$. A type-$\theta$ $M$-user facing an $A$-user obtains expected utility $\mu_A(R+b)+(1-\mu_A)S=P>o$, and therefore accepts. Two $A$-users also accept, since their continuation payoff is $P>o$. Thus, the adoption incentives of type $\theta$ are unchanged.

For type $\tau$, adopting $M$ yields a lower payoff against $M$-users, since $R<T$. Against $A$-users, the highest payoff from adopting $M$, allowing for refusal, is $\max\{\mu_A R+(1-\mu_A)S,o\}<P$. Hence type $\tau$ also has no incentive to deviate. The material fitness comparison is therefore the same as in the baseline defensive equilibrium, where $\Delta(\sigma, x)<0$.

Next, suppose $S<o<P$ and consider an equilibrium in which all type-$\theta$ individuals choose $M$, while all type-$\tau$ individuals choose $A$. Let $M$-users accept $M$-users and decline $A$-users, while $A$-users accept and defect against one another. Since every $A$-user is selfish, declining an $A$-user is optimal for an $M$-user, as $o>S$.

A type-$\theta$ individual obtains $x(R+b)+(1-x)o$ from choosing $M$ and $xo+(1-x)P$ from choosing $A$. Thus, choosing $M$ is optimal if and only if
\[
x\geq\frac{P-o}{R+b+P-2o}.
\]
Similarly, type $\tau$ prefers $A$ if and only if
\[
x\leq\frac{P-o}{R+P-2o}.
\]
The material fitness differential is
\[
\Delta(\sigma, x)=xR+(1-x)o-[xo+(1-x)P]<0
\]
whenever the second inequality is strict. The lower bound on $x$ is strictly below $\frac{P-S}{R+b-S}$ \cl{because $(P-S)(R+b+P-2o)-(P-o)(R+b-S)=(o-S)(R+b-P)+(P-o)(P-S)>0$.}

Finally, suppose $P\leq o<R$ and $x>\frac{o-S}{R+b-S}$.
Consider an equilibrium with $\Pr(M\mid\tau)=0$ and
\[
\Pr(M\mid\theta)
=1-\frac{(1-x)(o-S)}{x(R+b-o)}.
\]
The assumed restriction on $x$ ensures that this adoption probability lies strictly between zero and one. By Bayes' rule, the posterior probability that an $A$-user is type $\theta$ is
\[
\mu_A=\frac{o-S}{R+b-S}.
\]

Let $A$-users decline other $A$-users, but accept $M$-users. Type-$\theta$ $M$-users accept both technologies, while type-$\tau$ $M$-users decline $A$. Against $M$, type-$\theta$ $A$-users cooperate and type-$\tau$ $A$-users defect. The defensive continuation is played following two $A$ choices if interaction occurs.

A type-$\theta$ $M$-user is indifferent between accepting and declining an $A$-user, since
\[
\mu_A(R+b)+(1-\mu_A)S=o.
\]
A type-$\tau$ $M$-user instead prefers to decline an $A$-user. Two $A$-users decline one another because their defensive continuation payoff satisfies $P\leq o$.

Recall that $y=x\Pr(M\mid\theta)$ denotes the population share using $M$. A type-$\theta$ individual then obtains the same expected utility from either AI, which is $y(R+b)+(1-y)o$.

A type-$\tau$ individual obtains $yT+(1-y)o$ from choosing $A$, compared with $yR+(1-y)o$ from choosing $M$, and therefore strictly prefers $A$.

For material fitness, a type-$\theta$ $A$-user receives $yR+(1-y)o$. A type-$\theta$ $M$-user receives a weakly lower material payoff, since her expected material payoff against an $A$-user is $o-\mu_A b<o$. Hence
\[
F^\theta(\sigma, x)\leq yR+(1-y)o
<yT+(1-y)o=F^\tau(\sigma, x),
\]
where the strict inequality follows from $y>0$ and $T>R$.
\hfill$\square$
}

\bibliographystyle{ecta}
\bibliography{bib}

\end{document}